\documentclass[showkeys]{revtex4}
\usepackage{natbib}
\usepackage{amsmath}
\usepackage{amssymb}
\usepackage{amsfonts}
\usepackage{graphicx}
\usepackage{mathrsfs}
\usepackage{ulem} 

\begin{document}
\title[ ]{Clausius versus Sackur-Tetrode entropies}

\author{Thomas Oikonomou}
\email{thoikonomou@chem.demokritos.gr}

\affiliation{Department of Physics, Faculty of Science, Ege University, 35100 Izmir, Turkey}
\affiliation{Department of Physics, University of Crete, 71003 Heraklion, Hellas}
\affiliation{Institute of Physical Chemistry, National Center for Scientific Research
``Demokritos", 15310 Athens, Hellas}

\author{G. Baris Bagci}
\affiliation{Department of Physics, Faculty of Science, Ege University, 35100 Izmir, Turkey}



\begin{abstract}
Based on the property of extensivity (mathematically, homogeneity
of first degree), we derive in a mathematically consistent manner
the explicit expressions of the chemical potential $\mu$ and the
Clausius entropy $S$ for the case of monoatomic ideal gases in
open systems within phenomenological thermodynamics. Neither
information theoretic nor quantum mechanical statistical concepts
are invoked in this derivation. Considering a specific expression
of the constant term of $S$, the derived entropy coincides with
the Sackur-Tetrode entropy in the thermodynamic limit. We
demonstrate however, that the former limit is not contained in the
classical thermodynamic relations, implying that the usual
resolutions of Gibbs paradox do not succeed in bridging the gap
between the thermodynamics and statistical mechanics. We finally
consider the volume of the phase space as an entropic measure, albeit,
without invoking the thermodynamic limit to investigate its
relation to the thermodynamic equation of state and observables.
\end{abstract}
\eid{ }
\date{\today }
\startpage{1}
\endpage{1}
\maketitle

\section{Introduction}
\label{intro}

Gibbs paradox emerges when the same gas occupies two chambers of
identical volume $V$ separated by a partition \cite{Huang}. In
addition, both chambers contain equal number of particles $N$ and
possess the same total energy $U$, temperature $T$, and pressure
$P$. Then, the partition is suddenly removed so that the gases in
different chambers are allowed to mix. This is the famous Gibbs
scenario. The question is then what the change in entropy would be
having removed the partition?


The statistical entropy expressions found in the textbooks are shown to yield the entropy change $2k_{B}N\text{ln}2$ ($k_{B}$ denotes the Boltzmann constant) in the thermodynamic limit\footnote{The thermodynamic limit is the limit in which the number of particles and the volume go to infinity, while their ratio converges to a finite constant value while the energy per particle is also constant.} when applied to the mixing
of the gases upon removal of the partition. What is so particular about the value i.e., $2k_{B}N\text{ln}2$,
that one can now dub the case in hand as a paradox? The answer
lies not in statistical mechanics \textit{per se}, but its relation to
thermodynamics: the Gibbs scenario represents a reversible process
so that one beforehand knows that the entropy change must be equal
to zero according to the second law of thermodynamics, and
$2k_{B}N\text{ln}2$ is simply not equal to zero. This,
\textit{prima facie}, implies that the same process is reversible
for thermodynamics whereas it is irreversible when viewed in terms
of statistical mechanics. Since this cannot be true, one considers
it as paradox, and for historical reasons, dubs it as Gibbs
paradox \cite{Dieks2}.

However, note that thermodynamics is based on the Clausius definition of
entropy (this is why it is usually called the thermodynamic
entropy) and one assumes that the statistical definitions of
entropy are equivalent to the Clausius one so that they must yield
the same answer as the thermodynamic entropy \cite{Dieks2}. After
all, a main aim of statistical mechanics is that one can derive
the phenomenological macroscopic relations of thermodynamics from
a microscopic view. On the other hand, classical thermodynamics is
based on the assumption of extensivity, and therefore the
statistical entropy expressions too must preserve this assumption. The textbook
resolution of the Gibbs paradox explicitly emphasizes this point
so that all one has to do is to make statistical entropy
definitions somehow extensive \cite{Dieks2}. The usual way to do this is by dividing
the number of microstates, phase space volume or phase space
surface by $N!$ as a result of the (quantum-mechanical) indistinguishability of the particles, so that using the Stirling approximation, one can
now write down extensive statistical entropies and therefore Gibbs
paradox is resolved. This final extensive entropy expression is
called the Sackur-Tetrode (ST) entropy
\begin{eqnarray}\label{STE}
\nonumber
S_{\mathrm{ST}}(U,V,N)&=&\lim_{N\rightarrow\infty} k_{\mathrm{B}} \ln\bigg( \frac{\Phi}{h^{3N}N!}\bigg)=\lim_{N\rightarrow\infty} k_{\mathrm{B}} \ln\bigg( \frac{\Omega}{h^{3N}N!}\bigg)\\
&=& k_{\mathrm{B}}N \bigg[
    \ln\bigg(\frac{V}{N}\bigg) +\frac{3}{2}\ln\bigg(\frac{U}{N}\bigg) +\frac{5}{2}
    -\frac{3}{2}\ln\bigg(\frac{3h^2}{4\pi m}\bigg)
    \bigg]\,,
\end{eqnarray}
where $h$ is the Planck constant and $m$ denotes the mass of the
particle \cite{Huang}. The Sackur-Tetrode entropy has historically
been derived using quantum statistical mechanics for an ideal gas.
We also note that an information theoretical derivation of the
ST-entropy has been also demonstrated by A. Ben-Naim
\cite{Ben-Naim}, where the former function is the result of the
combination of the following four terms: positional uncertainty,
momenta uncertainty, quantum mechanical uncertainty principle and
the indistinguishability of the particles. In the same book one
may also find an alternative derivation of the factor $N!$, with
the Schr\"odinger equation being the point of departure.
$S_\mathrm{ST}$ is then expressed in terms of the de Broglie
wavelength.

The current discussions on Gibbs paradox generally revolves around
whether the factor $N!$ is justified due to the
indistinguishability of the particles (and therefore the
resolution of the paradox being quantum mechanical)
\cite{Allahverdyan}, or if it can be obtained classically
\cite{Cheng,Kampen,Swendsen2}. A possible third way out is to
argue that the very existence of the paradox points out to the
difference between the statistical and thermodynamical second laws \cite{Dieks2}.

It is essential to understand that the inclusion of the factor
$N!$ has the effect of making the statistical entropy expressions
extensive in the thermodynamic limit. Therefore, we aim to derive
the thermodynamical i.e., Clausius entropy for monoatomic ideal
gas for open systems by focusing on the property of extensivity,
albeit without invoking the thermodynamic limit. Then, we argue
the implications of this derivation for the ST-entropy and Gibbs
paradox. However, it is important to stress that the extensivity
property is not only important for the resolution of the Gibbs
paradox, but a general requirement for the homogeneous systems
(see pp. 971-972 in Ref. \cite{Callender}).

One might question the motivation for the derivation of the
thermodynamic entropy for monoatomic ideal gas, since many
textbooks treat it as trivial. However, one usually calculates it
by using $dU=TdS-PdV$ for a reversible process without taking the
term $\mu\, dN$ fully into account. Hence, one obtains, using also
the equation of state and explicit expression for the energy,
$S\left( U,V\right) =Nk_{\mathrm{B}}\left[ \frac{3}{2}\ln \left(
\frac{U}{N}\right) +\ln \left( V\right) \right]$ apart from some
irrelevant constant terms. The thermodynamic entropy obtained in
this manner is not a function of the number of particles despite
the appearance of $N$. Moreover, it is not extensive and therefore
not free of Gibbs paradox! Since we know that it has to be
extensive, the textbook account argues, we can divide the volume
$V$ by $N$, which is the thermodynamic \textit{ad hoc} equivalent
of dividing the number of microstates by $N!$ in statistical
mechanics (see p. 42 in \cite{Greiner} for an explicit statement
of this fact). Before proceeding further, it is noted that the
scope of the present paper is limited to homogeneous systems and
ideal gases with no surface effects. The systems with surface
effects, by their very nature, can be non-extensive and must be
treated as such.

The paper is organized as follows.
Section \ref{sec:HomoFun} reviews the basic relations
characterizing the homogenous functions of first degree i.e., the
extensivity. Having outlined the formalism of the former
functions, in Section \ref{sec:3Variables}, we apply it to
thermodynamic functions to derive the chemical potential and the
entropy expression for monoatomic ideal gases in open systems.
Section 4 considers the possible implications of the entropy
expression derived in the previous section for the ST-entropy and
Gibbs paradox.
Conclusions are presented in Section \ref{sec: Conclusions}.

\section{Thermodynamic extensivity}\label{sec:HomoFun}

A function $f:\mathbb{R}_+^q\rightarrow\mathbb{R}$ of class
$C^{n\geq2}$ is said to be homogeneous of $\tau$th degree ($\tau
d$) generally, when it satisfies the following relation
\begin{equation}\label{homokd}
f(\alpha\,\mathbf{x})=\alpha^\tau\,f(\mathbf{x})\,,
\end{equation}
where $\alpha\in\mathbb{R}_+$ and $\mathbf{x}\in\mathbb{R}_+^q$.
Then, homogeneity of degree one (1d) is characterized by Eq.
(\ref{homokd}) for $\tau=1$ i.e.,

\begin{equation}\label{NewEq:01}
f(\alpha\, \mathbf{x})=\alpha\, f(\mathbf{x})\,,
\end{equation}
which is also called extensivity \footnote{Although we use
1d-homogeneity and extensivity interchangeably, it is important to
keep in mind this rigorous mathematical definition which does not
depend on the thermodynamic limit at all. Moreover, this
definition ensures that one does not confuse the property of
extensivity with additivity (see for example Ref.
\cite{Touchette}).}. The differentiation of Eq. (\ref{NewEq:01})
with respect to $\alpha$ yields
\begin{equation}\label{NewEq:02}
\sum_{i=1}^{q} x_i\; \frac{\partial f(\alpha\,
\mathbf{x})}{\partial (\alpha\, x_i)}= f(\mathbf{x})\,.
\end{equation}
Eq. (\ref{NewEq:02}) is valid for any $\alpha\in\mathbb{R}_+$,
thus we can choose $\alpha=1$ without loss of generality.
We then obtain
\begin{equation}\label{NewEq:03}
f(\mathbf{x})=\sum_{i=1}^{q} x_i \, A_i(\mathbf{x})\,,
\end{equation}
where $A_i(\mathbf{x}):=\partial f(\mathbf{x})/\partial x_i$. The
above equation is referred to as \textit{Euler's 1d-homoge\-neous
function theorem}, stating that a 1dH-function $f(\mathbf{x})$ can
always be expressed in the form of Eq. (\ref{NewEq:03}).
It can easily be shown that the relation between Eq.
(\ref{NewEq:01}) and Eq. (\ref{NewEq:03}) is bijective.
From the latter equation we see that the functions
$A_i(\mathbf{x})$ are 0d-homogeneous (0dH) or equivalently
intensive, i.e., $A_i(\alpha\,\mathbf{x})=A_i(\mathbf{x})$.
Computing the partial derivative of $f(\mathbf{x})$ in Eq.
(\ref{NewEq:03}) with respect to $x_j$, we obtain
\begin{equation}\label{NewEq:04aa}
\frac{\partial f(\mathbf{x})}{\partial
x_j}=A_j(\mathbf{x})+\sum_{i=1}^{q}x_i\frac{\partial A_i(\mathbf{x})}{\partial x_j}\,,
\end{equation}
implying
\begin{equation}\label{NewEq:04ab}
\sum_{i=1}^{q}x_i\frac{\partial A_i(\mathbf{x})}{\partial x_j}=0\,.
\end{equation}
As can be seen, Eq. (\ref{NewEq:04ab}) is satisfied by any
1dH-function and thus it presents another characteristic relation
of the former class of functions. In fact, Eq. (\ref{NewEq:04ab})
presents a set of $q$ conditions, since $j=1,\ldots,q$.

Now, we want to prove the reverse, namely if Eq.
(\ref{NewEq:04ab}) is true, then $f(\mathbf{x})$ is 1dH.
Eq. (\ref{NewEq:04ab}) is valid for any translation
$\widetilde{f}(\mathbf{x})$ of $f(\mathbf{x})$, i.e.,
$\widetilde{f}(\mathbf{x})=f(\mathbf{x})+c$, where $c$ is a
constant, so that $\widetilde{A}_i(\mathbf{x}):=\partial
\widetilde{f}(\mathbf{x})/\partial x_i=\partial
f(\mathbf{x})/\partial x_i=A_i(\mathbf{x})$.
Multiplying Eq. (\ref{NewEq:04ab}) by $dx_j$ and then summing over
all $j$'s in the aforementioned equation yields
\begin{equation}\label{NewEq:05a}
\sum_{j=1}^{q}\sum_{i=1}^{q}x_i\frac{\partial
\widetilde{A}_i(\mathbf{x})}{\partial x_j}dx_j=
\sum_{i=1}^{q}\sum_{j=1}^{q}x_i\frac{\partial
\widetilde{A}_i(\mathbf{x})}{\partial x_j}dx_j
=\sum_{i=1}^{q}x_i\,d\widetilde{A}_i(\mathbf{x})=0\,.
\end{equation}
The change in the order of the summation in  Eq. (\ref{NewEq:05a})
can be performed, since
$\sum_{j=1}^{q}(\sum_{i=1}^{q}x_i\big[\partial
\widetilde{A}_i(\mathbf{x})/\partial
x_j\big]dx_j=0<\infty)<\infty$.
Adding the term $\sum_{i=1}^{q}\widetilde{A}_i(\mathbf{x})dx_i$ in
Eq. (\ref{NewEq:05a}), we obtain
\begin{eqnarray}\label{NewEq:05b}
\nonumber \sum_{i=1}^{q}\widetilde{A}_i(\mathbf{x})dx_i&=&
\sum_{i=1}^{q}\widetilde{A}_i(\mathbf{x})dx_i+\sum_{i=1}^{q}x_i\,d\widetilde{A}_i(\mathbf{x})
\quad\Rightarrow\\
\nonumber
d\widetilde{f}(\mathbf{x})&=&d\bigg(\sum_{i=1}^{q}x_i\widetilde{A}_i(\mathbf{x})\bigg)
\quad\Rightarrow\quad \widetilde{f}(\mathbf{x})=\sum_{i=1}^{q}x_i
\widetilde{A}_i(\mathbf{x})+a
\qquad\Rightarrow\\
f(\mathbf{x})&=&\sum_{i=1}^{q}x_i A_i(\mathbf{x})+(a-c)\,,
\end{eqnarray}
where $a$ is an integration constant.
We can always choose $c=a$, so that Eq. (\ref{NewEq:05b}) is
identified with Eq. (\ref{NewEq:03}).
Herewith, we have proven that the conditions in Eq.
(\ref{NewEq:04ab}) are necessary and sufficient conditions for
having 1d-homogeneity. Thus, they can be considered as an
alternative definition of 1dH-functions.
An important consequence of Eq. (\ref{NewEq:04ab}) (or
equivalently Eq. (\ref{NewEq:05a}) ) is that the differential of
the function $f(\mathbf{x})$ is exact
\begin{equation}
df(\mathbf{x})=\sum_{i=1}^{q}A_i(\mathbf{x})dx_i+\sum_{i=1}^{q}x_idA_i(\mathbf{x})
\overset{(\ref{NewEq:05a})}{=}\sum_{i=1}^{q}A_i(\mathbf{x})dx_i
=\sum_{i=1}^{q}\frac{\partial f(\mathbf{x})}{\partial x_i}dx_i\,,
\end{equation}
implying simply, that there are indeed functions satisfying Eq.
(\ref{NewEq:01}).

It is also worth remarking that Eq. (\ref{NewEq:04ab}) can be
rewritten as $\sum_{i=1}^{q}x_i\big[\partial
A_j(\mathbf{x})/\partial x_i\big]=0$, because of the equality
$\partial A_i(\mathbf{x})/\partial x_j=\partial
A_j(\mathbf{x})/\partial x_i$ which holds true for any
$C^{n\geq2}$-function. One can easily verify that the former
equation is the respective result of Eq. (\ref{NewEq:03}) for 0dH
functions. It becomes obvious then that Eq. (\ref{NewEq:04ab})
fully characterizes both the 1dH-function $f(\mathbf{x})$ and the
0dH-functions $A_i(\mathbf{x})$ simultaneously.

\section{Clausius entropy for open systems}
\label{sec:3Variables}

We now make use of the results presented in the previous section
on  1dH- and 0dH-functions to model thermodynamic quantities,
considering three independent variables, i.e, $U$, $V$ and $N$.
Further, we assume a 1dH-function $S=S(U,V,N)$ as the
thermodynamic entropy. Then, according to Eq. (\ref{NewEq:03}),
$S$ can be expressed as follows
\begin{equation}\label{EulerRel}
S(U,V,N)=\frac{1}{T(U,V,N)}U+\frac{P(U,V,N)}{T(U,V,N)}V-\frac{\mu(U,V,N)}{T(U,V,N)}N\,,
\end{equation}
where
\begin{eqnarray}\label{definitions}
\nonumber \frac{1}{T(U,V,N)}&:=&\frac{\partial S(U,V,N)}{\partial
U}\,,\qquad
\frac{P(U,V,N)}{T(U,V,N)}:=\frac{\partial S(U,V,N)}{\partial V}\,,\\
\frac{\mu(U,V,N)}{T(U,V,N)}&:=&-\frac{\partial S(U,V,N)}{\partial
N}\,.
\end{eqnarray}
In thermodynamics,  Eq. (\ref{EulerRel}) is referred to as the
\textit{Euler equation} \footnote{Note that the
thermodynamics is also applicable to the non-extensive systems so
that one can still define the thermodynamic potentials through Eq.
(12).}.
The left hand side of the definitions in Eq. (\ref{definitions})
are thermodynamic observables, where $\mu$ denotes the chemical
potential.
All thermodynamic observables must be 0dH-functions (intensive),
since otherwise the assumption of $S$ being 1dH (extensive) cannot
hold true.
As shown in the previous section, $S$ has to satisfy the
conditions in Eq. (\ref{NewEq:04ab}).
In the thermodynamic notation the former conditions take the form
\begin{subequations}\label{Constraints-1}
\begin{eqnarray}
U\frac{\partial}{\partial U}\frac{1}{T}+V\frac{\partial }{\partial U}\frac{P}{T}-N\frac{\partial }{\partial U}\frac{\mu}{T}&=0\,,\\
U\frac{\partial}{\partial V}\frac{1}{T}+V\frac{\partial }{\partial V}\frac{P}{T}-N\frac{\partial }{\partial V}\frac{\mu}{T}&=0\,,\\
U\frac{\partial}{\partial N}\frac{1}{T}+V\frac{\partial }{\partial
N}\frac{P}{T}-N\frac{\partial }{\partial N}\frac{\mu}{T}&=0\,.
\end{eqnarray}
\end{subequations}
If the 0dH-functions $P,T$ and $\mu$ satisfy Eq.
(\ref{Constraints-1}), then the differential of $S$ is exact
yielding the well-known first law of thermodynamics

\begin{equation}\label{FLT0}
T\,dS=dU+P\,dV-\mu\,dN\,,
\end{equation}
which is also the conservation of energy.\footnote{This
form of the first law implies, having $\delta Q=TdS$, reversible
processes.}

Considering the relations in Eq. (\ref{Constraints-1})
simultaneously, they can be written as
\begin{equation}\label{Gibbs-Duhem}
Ud\bigg(\frac{1}{T}\bigg)+Vd\bigg(\frac{P}{T}\bigg)-Nd\bigg(\frac{\mu}{T}\bigg)=0\,.
\end{equation}
We identify Eq. (\ref{Gibbs-Duhem}) with the well known
\textit{Gibbs-Duhem relation} \cite{Fliessbach}.

Let us now explore the conditions in Eq. (\ref{Constraints-1}) for
the case of an ideal monoatomic gas in an open system, $dN\neq0$.
The expressions of pressure and temperature are given by
\begin{equation}\label{TempPress}
P(U,V,N)=P(U,V)=\frac{2 U}{3 V}\,,\qquad T(U,V,N)=T(U,N)=\frac{2
U}{3 k_{\mathrm{B}} N}\,.
\end{equation}
The pressure in Eq. (\ref{TempPress}) has been derived within the
kinetic gas theory. The temperature in Eq. (\ref{TempPress}),
within classical thermodynamics, presents an \textit{ad hoc}
definition, being compatible with the equation of state for the
ideal gases, i.e., $PV=k_{\mathrm{B}}NT$. Indeed, combining $P$
and $T$ in Eq. (\ref{TempPress}) one obtains the former equation
of state.
Apparently, the functions $P$ and $T$ are 0dH.
In the mathematically trivial case of constant pressure and
temperature, $dP=0=dT\;\Rightarrow\; U\sim N$ and $V\sim N$, Eq.
(\ref{Constraints-1}) yields a constant chemical potential $\mu$.
In this case, the respective entropy function in Eq.
(\ref{EulerRel}) is 1dH and linear with respect to $N$,  i.e.,
$S(U,V,N)=S(N)\sim N$.
For the general case of varying pressure and temperature i.e.
$dP\neq0$ and $dT\neq0$, from Eq. (\ref{Constraints-1}), we read
\begin{subequations}\label{ChemPot}
\begin{eqnarray}
\label{ChemPot1}\frac{\partial}{\partial U}\mu(U,V,N)&=&\frac{1}{U}\mu(U,V,N)-\frac{1}{N}\,,\\
\label{ChemPot2}\frac{\partial}{\partial V}\mu(U,V,N)&=&-\frac{2U}{3VN}\,,\\
\label{ChemPot3}\frac{\partial}{\partial
N}\mu(U,V,N)&=&-\frac{1}{N}\mu(U,V,N)+\frac{5U}{3N^2}\,,
\end{eqnarray}
\end{subequations}
respectively.
Solving this system of differential equations, we are able to
determine the expression of the chemical potential compatible with
Eqs. (\ref{FLT0}) and (\ref{TempPress}).
The aforementioned system can be solved stepwise. We first solve
Eq. (\ref{ChemPot1}). The result will depend on an integration
constant $Z(V,N)$, with respect to $U$.  Then, plugging the former
result into Eq. (\ref{ChemPot2}), $Z(V,N)$ reduces to an
integration constant $Z'(N)$. Finally, following the same
procedure with Eq. (\ref{ChemPot3}), $Z'(N)$ reduces to a
constant, which cannot be further reduced.
Thus, for varying pressure and temperature, the monoatomic ideal
gas chemical potential is determined apart from a constant $\xi$
as
\begin{equation}\label{ChemPotNew}
\mu(U,V,N)=\frac{U}{N}\bigg[\xi+
\ln\bigg(\frac{N}{U}\bigg)+\frac{2}{3}\ln\bigg(\frac{N}{V}\bigg)\bigg]\,.
\end{equation}
As expected, the $\mu$-function is 0dH.
The respective entropy function $S$ in Eq. (\ref{EulerRel}) is
then 1dH, given as
\begin{equation}\label{SackurTetrode}
S(U,V,N)=k_{\mathrm{B}} N\bigg[
        \ln\bigg(\frac{V}{N}\bigg) + \frac{3}{2}\ln\bigg(\frac{U}{N}\bigg) + \frac{5}{2}-\frac{3}{2}\xi\bigg]\,.
\end{equation}
Studying the validity of $S$, we observe that  the function $S$ in
Eq. (\ref{SackurTetrode}) can be derived by means of the
1d-homogeneity formalism when $P,T,\mu\neq\{0,const.,\pm\infty\}$,
beforehand, and when $dU\neq0$, $dV\neq0$ and $dN\neq0$,
simultaneously. If one of the former differentials is equal to
zero, implying that the respective variable is a constant, then
the 1d-homogeneity formalism is not applicable, since in this case
not all of the partial derivatives of $S$, i.e., $1/T$, $P/T$ or
$\mu/T$ are 0dH.

Comparing now the Clausius ideal gas entropy $S$ in Eq.
(\ref{SackurTetrode}) with the Sackur-Tetrode entropy
$S_\mathrm{ST}$ in Eq. (\ref{STE}), we observe that they are the
same when $\xi$ takes the value $\xi=\ln\big(\frac{3h^2}{4\pi
m}\big)$.
This specific expression of $\xi$ cannot be obtained by virtue of
the 1d-homogeneity analysis but only within quantum statistical
mechanics.

Before closing this section, we note that Eq. (19) can also be
found in \cite{Callen} where the thermodynamic extensivity is
implicitly assumed, and not focused in detail.

\section{On Gibbs Paradox}\label{sec: Discussions}

We now explore the implications of the derived thermodynamic
results on the Gibbs paradox. If one compares Eq.
(\ref{SackurTetrode}) with the usual expression found in the
textbooks i.e., Eq. (\ref{STE}), one is inclined to consider that
the only difference lies in the additive term including the Planck
constant. This would still not be a major difference in the sense
that this additive term would be washed out when one considers the
entropy change, so that it cannot be important for the resolution
in Gibbs paradox. In other words, quantum mechanics might be
interesting to fully determine the entropy expression, but it is
irrelevant considering the entropy changes and hence for the Gibbs
paradox.

On the other hand, this is not the whole story, since Eq.
(\ref{SackurTetrode}) is derived by assuming a nonconstant
temperature (making its direct application to Gibbs paradox
untenable, since Gibbs paradox assumes constant temperature)
contrary to Eq. (\ref{STE}). As a matter of fact, the
thermodynamic formulation of the previous section indicates that
for constant $\left( T,P,N\right)$ just as required by Gibbs
paradox for each compartment, we have $S(U,V,N)=S(N)\sim N$, since
one then concludes that $\frac{U}{V}$ and $\frac{U}{N}$ are
constant (so is then $\frac{V}{N}$ ) as can be seen from Eq. (16).
Note that this information on the entropy expression is enough to
understand that the thermodynamic entropy change is zero for the
Gibbs paradox, indicating reversibility of the
process.\footnote{This is so, since the total entropy before the
removal of the partition i.e., $S(N)\sim \left(
N_{1}+N_{2}\right)$, is unchanged after the removal of the
partition, where the lower indices denote two partitions.} This
proportionality i.e., $S(N)\sim N$, is indeed what one concludes
also from Eq. (\ref{STE}), considering that it is derived under
the assumption of fixed $T$ and $N$, since one then has $U\sim N$
and $V\sim N$. Therefore, the terms $\frac{U}{N}$ and
$\frac{V}{N}$ can be considered as a constant, hence the sole term
of importance is proportionality to $N$, namely,
$S_{\mathrm{ST}}\sim k_{\mathrm{B}}N$. It is worth remarking that
the rigorous thermodynamic treatment in the previous section is
far more general than the usual textbook treatments so that Eq.
(\ref{SackurTetrode}) is more general than Eq. (\ref{STE}) in its
scope of applications if one does not consider the
quantum-mechanical additive term.

Another important difference is concerning the thermodynamic
limit. The present thermodynamic treatment shows no need for the
thermodynamic limit, whereas Eq. (\ref{STE}) is derived in the
thermodynamic limit as all other resolutions of Gibbs paradox. To
shed light on this issue , namely whether phenomenological
thermodynamics implicitly contains  the thermodynamic limit,  or
equivalently whether $S$ and $S_\mathrm{ST}$ are identical, we
consider the general solution (gs) of the first law of
thermodynamics in Eq. (\ref{FLT0}), regarding the ideal gas
entropy function.

%
%
The latter is given as in Eq. (\ref{SackurTetrode}), with the
$\xi$-term being a function of $N$, i.e., $\xi\rightarrow
\Theta(N)$, namely
\begin{equation}\label{GenSol}
S_{\mathrm{gs}}(U,V,N)= k_\mathrm{B}N\bigg[
        \ln\bigg(\frac{V}{N}\bigg) + \frac{3}{2}\ln\bigg(\frac{U}{N}\bigg) +
        \frac{5}{2}-\frac{3}{2}\,\Theta(N)\bigg]\,.
\end{equation}
In this general case, the chemical potential,
$\mu_\mathrm{gs}=-T(\partial S_\mathrm{gs}/\partial N)$, may
generally not preserve the  0dH property.
From  Eqs. (\ref{STE}) and (\ref{GenSol}), we observe that in the
case of ST-entropy the $\Theta$-function in finite systems is
inhomogeneous and has the explicit expression
\begin{equation}\label{STE-Theta}
\Theta_\mathrm{ST}(N)=\xi+\frac{2}{3}\Big[1-\ln(N)+\frac{\ln(N!)}{N}\Big]\,,
\end{equation}
satisfying the
$\lim_{N\rightarrow\infty}\Theta_\mathrm{ST}(N)=\xi$ limit.
%
%
Let us then explore the most general expression of $\Theta(N)$
compatible with 1d-homogeneity, $S_\mathrm{gs}\rightarrow
S^\mathrm{1dH}_\mathrm{gs}$ (and $\mu_\mathrm{gs}\rightarrow
\mu^\mathrm{0dH}_\mathrm{gs}$). Therefore, we compute again the
conditions in Eq. (\ref{Constraints-1}), in terms of
$S_\mathrm{gs}$ now. From the first two conditions we do not
obtain any new information, yet the last one yields
\begin{equation}\label{important1}
N\,\frac{\partial^2\Theta(N)}{\partial N^2}+2\, \frac{\partial
\Theta(N)}{\partial N}=0\qquad\Rightarrow\qquad
\Theta(N)=a_1-\frac{a_2}{N}\,,
\end{equation}
where $a_1$ and $a_2$ are integration constants.
Plugging the result in Eq. (\ref{important1}) into Eq.
(\ref{GenSol}), we obtain
\begin{eqnarray}\label{entfinal}
\nonumber
S^{\mathrm{1dH}}_{\mathrm{gs}}(U,V,N)&=& \frac{1}{T(U,V,N)}U+\frac{P(U,V,N)}{T(U,V,N)}V-\frac{\mu^\mathrm{0dH}_\mathrm{gs}(U,V,N)}{T(U,V,N)}N + \frac{3}{2} k_\mathrm{B} a_2\\
&=& k_\mathrm{B} N \bigg[\ln\bigg(\frac{V}{N}\bigg) + \frac{3}{2}
\ln\bigg(\frac{U}{N}\bigg) + \frac{5}{2} - \frac{3}{2}a_1 \bigg] +
\frac{3}{2}k_\mathrm{B}a_2\,,
\end{eqnarray}
where $\mu^\mathrm{0dH}_\mathrm{gs}=\mu$ in Eq. (\ref{ChemPotNew})
with $a_1\equiv \xi$.
Comparing Eqs. (\ref{SackurTetrode}) and (\ref{entfinal}), it
becomes obvious that $S^\mathrm{1dH}_{\mathrm{gs}}$ contains an
extra additive constant. However, in Eq. (\ref{NewEq:05b}) we
showed that such a constant can always be transformed away. Thus,
the $\frac{3}{2}k_\mathrm{B}a_2$-term has neither physical nor
mathematical impact on $S^\mathrm{1dH}_{\mathrm{gs}}$ so that the
former is identified with $S$ in Eq. (\ref{SackurTetrode}).
Eq. (\ref{important1}) then unveils that the thermodynamic limit
is \textit{not} included in the phenomenological thermodynamic
equations. If it would be included, then the $a_2$-term would not
appear, $\lim_{N\rightarrow\infty}a_2/N=0$.
Accordingly, the Clausius 1dH entropy function $S$ in Eq.
(\ref{SackurTetrode}), derived by means of purely thermodynamic
relations, is distinct from the 1dH ST-entropy, derived within
statistical mechanics in Eq. (\ref{STE}).

Although we have discussed the discrepancy related to the
thermodynamic limit above only for the case of ST-entropy, the
main \textit{morale} can be readily applied to all textbook
attempts of resolving Gibbs paradox, since all of them, in
bridging the gap between thermodynamics and statistical mechanics,
make use of the thermodynamic limit argument. However, it is
evident from the discussions so far that thermodynamics does not
invoke thermodynamic limit at all, whereas statistical mechanics
depends on it to yield the same results obtained in
thermodynamics. In this sense, the division by $N!$ in Eq.
(\ref{STE}) seems to be \textit{ad hoc} even though it is said to
be justified by arguments regarding classical or quantum
mechanical statistical mechanics, since the thermodynamics by
itself does not invoke any such limit.

Turning our attention to the statistical mechanics now, we can
analyze whether a statistical description without thermodynamic
limit is possible at all. If possible, this description must be
valid independent of the number of degrees of freedom. Therefore, a thermodynamically consistent choice among the statistical definitions of entropy is
in order. For this purpose, we consider the entropy in terms of the phase space
volume $S_{\Phi}$ as an example. The volume of the phase space
for the ideal gas reads \citep{Huang}

\begin{equation}\label{volumeone}
\Phi (U,V)=C_{3N}\left[\frac{V}{h^{3}}(2mU)^{3/2}\right]^{N},
\end{equation}

where $C_{3N}=\frac{\pi ^{3N/2}}{\Gamma (\frac{3N}{2}+1)}$. The
$\Gamma(x)$ in the denominator represents the gamma function with
the argument $x$. In order to obtain the volume entropy, it
suffices to take the logarithm of the phase space volume, and
multiply it by the Boltzmann constant $k_{\mathrm{B}}$ so that

\begin{equation}\label{volumetwo}
S_{\Phi }(U,V)=k_{\mathrm{B}}\ln(A)+Nk_{\mathrm{B}}\ln(V)+\frac{3Nk_{\mathrm{B}}}{2}\ln(U),
\end{equation}

where $A$ is a constant depending on $h$, $m$ and $N$. The
thermodynamics of the system can be calculated by taking partial
derivative of the volume entropy with respect to extensive
variables i.e., energy $U$ and volume $V$. The first partial
derivative evidently provides the expression for the temperature
i.e.,

\begin{equation}\label{volumethree}
T_\Phi = \bigg(\frac{\partial S_{\Phi }(U,V)}{\partial
U}\bigg)^{-1}=\frac{2U}{3Nk_{\mathrm{B}}}.
\end{equation}

The subscript $\Phi$ indicates that the quantities are calculated
by using the volume entropy. The partial derivative of the volume
entropy with respect to extensive variable $V$ provides the
equation of state for the ideal gas

\begin{equation}\label{volumefour}
\frac{P_{\Phi}}{T_{\Phi}} = \frac{\partial  S_{\Phi
}(U,V)}{\partial V}=\frac{Nk_{\mathrm{B}}}{V}.
\end{equation}

It is very important to realize that all the equations above, the
equation of state and the temperature for the ideal gas, are
obtained in terms of the volume entropy without invoking the
thermodynamic limit. Moreover, combining Eqs. (\ref{volumethree})
and (\ref{volumefour}), we can also obtain

\begin{equation}\label{volumefive}
P_{\Phi}(U,V)=\frac{2 U}{3 V}.
\end{equation}

The inspection of Eqs. (\ref{volumethree}) and (\ref{volumefive})
shows that they are exactly the same as Eq. (\ref{TempPress}).
This shows that, contrary to the widely held view, the statistical
treatments do not \textit{only} have to be valid for large number
of particles although the same expressions are obtained in
phenomenological thermodynamics for macroscopic systems (or
through kinetic theory). We also note that the equipartition
theorem, $\frac{kT}{2}=\left\langle
\frac{p^{2}}{2m}\right\rangle$, which defines the kinetic
temperature in terms of the microcanonical probability density
average is also valid independently of the number of degrees of
freedom if one uses the volume entropy $S_{\Phi}$
\citep{Dunkel,Campisi}.

We emphasize again that all these nice results are obtained
without the thermodynamic limit. Therefore, it is natural to ask
where the thermodynamic limit is needed if it is needed at all.
Despite the results so far, it is easy to see that even the volume
entropy falls short of being the thermodynamic Clausius entropy,
since it is not extensive as can be seen from Eq.
(\ref{volumetwo}). One must instead have a term $\frac{V}{N}$ so
that one can also resolve the Gibbs paradox in terms of a
statistical entropy expression i.e., the volume entropy in this
case. Due to Eq. (\ref{volumeone}), one sees that dividing the
volume $V$ by $N$ is tantamount to dividing the phase space volume
$\Phi$ by $N^{N}$. Therefore, if thermodynamics does not include
the thermodynamic limit, the equivalence between the thermodynamic
and statistical entropies requires the division of the phase space
volume by $N^{N}$. However, assuming that the statistical physics
is valid for large number of particles only, the term $N^{N}$ is
approximated by the Stirling approximation, namely, $N^{N}\sim
e^{N}N!$. Then, all the previous thermodynamic equalities such as
the definition of temperature and ideal gas state equation remain
valid also within Stirling approximation, since they have been
valid for any number of particles to begin with. In addition to
this, one resolves the Gibbs paradox in the thermodynamic limit by
relying on the Stirling approximation.

On the other hand, the thermodynamic entropy is extensive without
thermodynamic limit as shown in the previous section, since the
definition of extensivity in thermodynamics is rigorously defined
as a mathematical property without a limiting procedure.
Therefore, a true equivalence can only be constructed by finding
a way to divide the phase space volume by $N^{N}$ as a result of
mixing in Gibbs scenario without making use of the thermodynamic
limit. Otherwise, one either defines a new reduced statistical
entropy, and maintains that it is the same as the thermodynamical
one \citep{Cheng} only in the thermodynamic limit, or defends the
view that the statistical mechanical second law is different than
the thermodynamical one \citep{Dieks2}.

\section{Conclusions}\label{sec: Conclusions}

We have shown that the Clausius i.e. thermodynamic entropy of a
monoatomic ideal gas in its most general form (including open
systems) can be obtained by taking the property of extensivity as
the point of departure. This thermodynamic entropy expression
seems not to differ from the statistical ST-entropy apart from an
additive quantum-mechanical term, which can be neglected when one
considers the change in entropies as required by the second law of
thermodynamics. Therefore, we conclude that the quantum mechanical
arguments should not play a decisive role in the resolution of
Gibbs paradox, since one considers the thermodynamic second law in
its resolution, rendering the constant additive terms unimportant.
Note that the irrelevance of the quantum mechanics for the Gibbs
paradox can also be argued by observing that the narrow wave
packets too can follow almost classical trajectories, and hence
they can be considered to be distinguishable (see Ref.
\citep{Dieks2} for more on this issue).

Another important difference between thermodynamics and
statistical mechanics concerning the Gibbs paradox is that the
former does not invoke the thermodynamic limit whereas the latter
does. In other words, even though one can justify the usual
division by $N!$ in statistical mechanics (be it quantum or
classical arguments), this does not warrant a resolution of Gibbs
paradox, if the paradox is accepted to stem from the comparison
between the thermodynamic and statistical second laws. The reason
is that the thermodynamic entropy conforms to the second law for
reversible processes independent of the number of degrees of
freedom, whereas the statistical entropy measure succeeds at this
only in the thermodynamic limit (note that even ST-entropy is
obtained in statistical mechanics by appeal to the thermodynamic
limit). An exact equivalence between the thermodynamics and
statistical mechanics concerning the Gibbs paradox can be achieved
fully either through dividing the volume $V$ by $N$, or
equivalently, the phase space volume $\Phi$ by $N^{N}$, not by the
term $N!$ which requires the thermodynamic limit.

Finally, we remarked that some important thermodynamical equations
such as the definition of temperature and the equation of state
for the ideal gas can be derived in statistical mechanics just as
in thermodynamics i.e. without invoking the thermodynamic limit by
adopting the volume entropy $S_{\Phi}$. However, even in this
case, one is forced to find a justification for dividing the
volume $V$ by $N$ (or equivalently, the phase space volume $\Phi$
by $N^{N}$), or accept that the statistical entropy expression is
not the same as the thermodynamic Clausius definition.

Although we have mainly focused on the mathematical structure of
the thermodynamics and its relation to the statistical mechanics
with an emphasis on the Gibbs paradox in particular, it is worth
noting that the thermodynamics is phenomenological and therefore
of experimental nature. For this aspect of the thermodynamics and
its possible relation to the resolution of the Gibbs paradox, we
refer the interested reader to Ref. \citep{Corti}.

\section*{Acknowledgments}
%
We thank Sumiyoshi Abe for fruitful discussions and Joshua T.
Berryman for corrections in English. T.O. acknowledges
partial support by the THALES Project MACOMSYS, funded
by the ESPA Program of the Ministry of Education of Hellas.

\bibliographystyle{model1a-num-names}

\end{document}